# Influence of Similar Atom Substitution on Glass Formation in (La-Ce)-Al-Co Bulk Metallic Glasses

Ran Li, Shujie Pang, Chaoli Ma, Tao Zhang *

Department of Materials Science and Engineering, Beijing University of Aeronautics and Astronautics, Beijing 100083, China

**Abstract**

The glass-formation range of bulk metallic glasses (BMGs) based on lanthanum and cerium was pinpointed in La-Al-Co, Ce-Al-Co and pseudo-ternary (La-Ce)-Al-Co system respectively by copper mold casting. Through the stepwise substitution of La for solvent Ce in $(La_xCe_{1-x})_{65}Al_{10}Co_{25}$ alloys (0<x<1), the fully glassy rods of the $(La_{0.7}Ce_{0.3})_{65}Al_{10}Co_{25}$ alloy can be successfully produced up to 25 mm in diameter by tilt-pour casting. Comparing with the glass-forming ability (GFA) of single-lanthanide based alloys, $La_{65}Al_{10}Co_{25}$ and $Ce_{65}Al_{10}Co_{25}$, the coexistence of La and Ce with similar atomic size and various valence electronic structure can obviously improve the GFA of $(La_xCe_{1-x})_{65}Al_{10}Co_{25}$ BMGs, which can't be explained by the former GFA criteria for BMGs, e.g. atomic size mismatch and negative heats of mixing. A thermodynamic model was proposed to evaluate this substitution effect, which gives a reasonable explanation for the obvious improvement of GFA induced by the coexistence of similar atoms.

* Corresponding author: Tao Zhang
  E-mail: zhangtao@buaa.edu.cn

# 1. Introduction

Since a ternary La-Al-Ni bulk metallic glass (BMG) of 2.5 mm in diameter was successfully produced through adding Ni to a known binary La-Al amorphous alloy by copper mold casting in 1989 [1], many new ternary alloys with high glass-forming ability (GFA), such as Ln-Al-TM (Ln = lanthanide metal; TM = Co, Ni, Cu), Zr-Al-TM, Pd-Ni-P, Pt-Ni-P, and Mg-Cu-Y etc. [2-6], represented in the component formula as A-B-C, were discovered through the elemental addition in former binary glassy alloys in following years. Further investigations indicated that A, B and C can be classified into large, middle and small atoms according to their significantly different atomic radii [7]. It could improve the atomic packing efficiency in supercooled liquid and increase the difficulty of crystallization, which results in a higher GFA comparing with the former binary amorphous alloys [7-10]. Besides the improvement of rapid solidification technique, component diversification by multi-element substitution in a known binary or ternary alloy system to obtain multicomponent alloy systems, well known as "confusion principle" [11], is recognized as one of the most effective ways to improve the GFA of metallic glasses. However, the search of BMGs with high GFA in multicomponent alloy systems is very tedious and time consuming, and how to choose substituting elements to improve the GFA of the resulting alloy is still a confused problem.

Some empirical rules were proposed to achieve BMGs with high GFA in multicomponent systems as follows [7,12,13]: (1) the multicomponent system should consist of more than three element, and the main constituent elements should be satisfied with significant difference in atomic sizes above 12 % and negative heats of mixing with other constituent elements; (2) considering of the reduced glass-transition temperature, $T_{rg}$, a deep eutectic composition caused by frustrating crystallization should be chosen in the multicomponent system with chemical and topological differences of constituent elements. These criteria suggest that choosing the dissimilar atoms with obvious different atomic size with other constituent elements as substituting elements could be an effective way to improve the GFA of multicomponent alloys.

In recent years, many BMGs with super high GFA, e.g. Mg-, Ca-, Ti-, Zr-, Fe-, Co-, Cu-, Y-, Pd-, Pt-, Ln- (Ln=lanthanide metals) based BMGs [14-25], were successfully produced through multi-component substitution. (In this paper, we only consider the partial substitution with atomic content above 5 %, and the role of minor alloying additions will not be discussed [26].) We can class the substitution method as follows. The general substitution method is based on the above empirical criterion of atomic size mismatch. It can be named as "column substitution", because elements in the same group with A, B or C in the Periodic Table of the Elements (PTE) are chosen as substituting elements, which usually have distinct atomic sizes but similar valence electronic structure to the substituted one. Many multicomponent BMGs with higher GFA were reported through the partial column substitution, e.g. partially substituting Ag for solute and solvent Cu in Mg-(Cu-Ag)-Y and (Cu-Ag)-(Zr-Ti) respectively[14,20], Mg for solvent Ca in (Ca-Mg)-Ni [15], Sc for solvent Y in (Y-Sc)-Al-Co [16] and Ti for solute Zr in Cu-(Zr-Ti) [27]. Another method of element substitution, which is much less mentioned, can be named as "row substitution". By using the neighboring elements in the same period with A, B or C in PTE as the substituting components, which have similar atomic sizes but various valence electronic structures to the substituted one, the GFA can also be improved effectively. The representative samples are the partial substitution of Cu for solute Ni in Zr-Al-(Ni-Cu), Pt-(Ni-Cu)-P, Pd-(Ni-Cu)-P and La-Al-(Ni-Cu) [17,22,23,28], Co for solvent Fe in (Fe-Co)-(Si-B)-Nb [18], Fe for solvent Ni in (Ni-Fe)-(Si-B)-Nb [29] and Pt for solvent Pd in (Pd-Pt)-Cu-P [30], in which GFA can be increased remarkably by the row substitution. Recently, we reported that the GFA can be evidently improved by substituting multi-lanthanide atoms for the single-lanthanide solvent atom in (Ce-La-Pr-Nd)-Al-Co system [31], which is another example of row substitution. Up to now, little experimental and theoretical analysis has been performed on this substituting method, because the neighboring element in the same period has similar atomic radius to the substituted one, causing almost no change in the magnitude of atomic size mismatch in the alloy systems.

In this work, we developed a pseudo-ternary alloy, (La-Ce)-Al-Co, with superior GFA by the row substitution. Through the partial substitution of La for Ce in $(La_xCe_{1-x})_{65}Al_{10}Co_{25}$ alloy system, the fully glassy rod of 25 mm in diameter was produced by tilt-pour casting. The influences of this row substitution on the glass-forming ability, crystallization process and behaviors of melting and solidification were evaluated in detail. A thermodynamic model was proposed to give a possible explanation to understand this effect of the row substitution with similar atoms on the glass formation.

## 2. Experimental Procedure

All ingots of (La-Ce)-Al-Co, La-Al-Co and Ce-Al-Co alloys with nominal composition were prepared by arc melting a mixture of pure Co (99.9 mass%), Al (99.99 mass%), La and Ce (above 99.5 mass%) in a highly pure argon atmosphere. For the smaller rod-shaped sample (≤ 12 mm in diameter), the ingot was remelted in a quartz tube using an induction heating coil in a highly pure argon atmosphere, and then injected into a copper mold through a nozzle using a highly pure argon atmosphere at 0.2 atm. pressure to produce glassy rods. For the larger rod-shaped sample (≥ 15 mm in diameter), the ingot was remelted in a quartz cup using a tilting induction furnace and then the molten alloy was poured into a copper mold in a highly pure argon atmosphere. Cross sections of as-cast rods were examined by X-ray diffraction (XRD) using a Bruker AXS D8 X-ray diffractometry with Cu-Kα radiation at a scanning rate of 1 degree/minute to ensure the phase structure. The thermal stability of the glassy samples was evaluated by a NETZSCH DSC 404 C Differential Scanning Calorimeter (DSC) at a heating rate of 0.33 K/s in a flowing purified argon atmosphere. Melting and solidification behaviors of these alloys were also characterized by DSC at a heating and cooling rate of 0.33 K/s. For density measurement, the bulk glassy rods with the same diameter of 2 mm were measured by the liquid displacement method using a density determination kit (YDK01-0D, Sartorius AG) in 1,1,2,2-tetrabromoethane with an accuracy within 0.5 %.

## 3. Results and Discussion

3.1 Glass Formation of (La-Ce)-Al-Co system

In order to study the effect of row substitution with similar atoms and pinpoint the BMGs with superior GFA in the pseudo-ternary (La-Ce)-Al-Co system, a good glassy former must be found out so that the further substitution can be processed. We adopted the equiatomic ratio between La and Ce to search the composition map of $(La_{0.5}Ce_{0.5})$-Al-Co system based on the latest research [32].

Figure 1 shows the composition range for glass formation of $(La_{0.5}Ce_{0.5})$-Al-Co BMGs. It indicates that the BMGs with the glassy critical diameter ($d_c$, $d_c$ means the maximal diameter in which the fully glassy rod can be produced successfully) not less than 2 mm can be produced in a large composition range (the blue marks), and the fully glassy samples of at least 12 mm in diameter can also be produced in a local range (the red marks), confirmed by the smooth XRD patterns in Fig. 2. We evaluated the melting characteristics of the alloys in the composition map by the DSC analysis, and the 3D contour map of liquidus surface was plotted, as shown in Fig. 3. Thermal parameters of $(La_{0.5}Ce_{0.5})$-Al-Co BMGs, e.g. the glass transition temperature ($T_g$), the onset temperature of crystallization ($T_x$), the melting temperature ($T_m$), liquidus temperature during heating process ($T_l$), the supercooled liquid region ($\Delta T_x = T_x - T_g$), $T_{rg}$ ($T_{rg} = T_g / T_l$) and $\gamma$ ($\gamma = T_x / (T_g + T_l)$) [33], are shown in Table I. We noticed that the alloys with $d_c$ above 12 mm lie on relative lower liquidus surface, nevertheless deviate from the eutectic composition. Although the alloy composition with the largest $T_{rg}$ is $(La_{0.5}Ce_{0.5})_{70}Al_{10}Co_{20}$, the alloys having the highest GFA are $(La_{0.5}Ce_{0.5})_{70}Al_{15}Co_{15}$, $(La_{0.5}Ce_{0.5})_{65}Al_{10}Co_{25}$ and $(La_{0.5}Ce_{0.5})_{65}Al_{15}Co_{20}$ according to their values of $d_c$. The relationship between the GFA and the $\Delta T_x$ or $\gamma$ is also incoherent. The possible explanations have been mentioned in some literatures [24,34]. The BMGs with obvious supercooled liquid region $\Delta T_x$ over 65 K (the maximum reaches 86 K) can be found in a large composition range of $(La_{0.5}Ce_{0.5})$-Al-Co system, which is hopeful to be a further superplastic application as mentioned in references [35,36].

Considering the melting behavior and GFA, we chose the (La-Ce)$_{65}$Al$_{10}$Co$_{25}$ alloy as an appropriate glassy former to further study the influence of partial row substitution on the GFA of (La$_x$Ce$_{1-x}$)$_{65}$Al$_{10}$Co$_{25}$.

3.2 Substitution Effect on Glass Formation of (La$_x$Ce$_{1-x}$)$_{65}$Al$_{10}$Co$_{25}$

In order to study the substitution effect of similar atoms of La and Ce on the GFA of resulting alloys, firstly, we produced the glassy rods of (La$_x$Ce$_{1-x}$)$_{65}$Al$_{10}$Co$_{25}$ alloys ($0 < x < 1$) of 2 mm in diameter through the row substitution of La for solvent Ce. The glass transition, crystallization and melting behaviors of these alloys were studied by the DSC at a heating and cooling rate of 0.33 K/s, and the thermal parameters, e.g. $T_g$, $T_x$, $T_m$, $T_l$, $T_{l'}$ (the liquidus temperature during cooling process), $\Delta T_x$, $\Delta T_l$ ($\Delta T_l = T_l - T_{l'}$, $\Delta T_l$ is the nominal supercooled degree at a certain same heating and cooling rate), $T_g/T_m$, $T_g/T_l$, $T_g/T_{l'}$ and $\gamma$, as well as the density of glassy samples, were listed in Table II. Some other La- and Ce-based BMGs were also included for comparison [24,25]. Figure 4 (a) shows the DSC curves of the BMGs of (La$_x$Ce$_{1-x}$)$_{65}$Al$_{10}$Co$_{25}$ alloys in the heating process. With the increase of the substituting content of La, the BMGs exhibit higher $T_g$ and $T_x$. The substitution of La for solvent Ce in the (La$_x$Ce$_{1-x}$)$_{65}$Al$_{10}$Co$_{25}$ alloys causes the change of the number of the main characteristic exothermic peaks for the crystallization process. When the $x$ ranges from 0.5 to 0.8, the number of main crystalline peaks (n) reaches to the maximum, four peaks. While more solvent Ce atom was substituted by La ($x = 0.9$), the n decreases to 3. Moreover, it has been mentioned that the n of La$_{65}$Al$_{10}$Co$_{25}$ and Ce$_{65}$Al$_{10}$Co$_{25}$ is only 2 [31]. It indicates that the coexistence of La and Ce in the (La$_x$Ce$_{1-x}$)$_{65}$Al$_{10}$Co$_{25}$ BMGs with an appropriate ratio may increase the complexity of crystallization for the transformation from the metastable undercooled liquid state to the complete crystalline compound during the heating process. For the melting behavior, with the increase of the substituting content of La, the change of $T_m$ of the resulting alloys is indistinctive. So does the change of $T_l$ when the substituting content $x$ is below 0.5. When $x$ is above 0.5, the $T_l$ increases obviously from 776 K for $x = 0.5$ to 937 K for $x =$ 0.9. Figure 4 (b) shows the DSC cooling curves of $(La_xCe_{1-x})_{65}Al_{10}Co_{25}$ alloys ($0 < x < 1$). The change of $T_{l'}$ is similar with that of $T_l$.

Furthermore, we tried to find out the $d_c$ for the $(La_xCe_{1-x})_{65}Al_{10}Co_{25}$ BMGs in order to evaluate the influence on the GFA by the substitution with similar atoms. Figure 5 shows the XRD patterns for the $(La_xCe_{1-x})_{65}Al_{10}Co_{25}$ glassy rods with the maximal size in diameter, which confirms the fully amorphous structure for the samples. Through the row substitution with similar atoms, we obtained the BMGs with superior GFA in the alloy composition of $(La_{0.7}Ce_{0.3})_{65}Al_{10}Co_{25}$, which can successfully forms a fully glassy rod of 25mm in diameter. The outer shape and surface appearance of the as-cast $(La_{0.7}Ce_{0.3})_{65}Co_{25}Al_{10}$ and $(La_{0.6}Ce_{0.4})_{65}Co_{25}Al_{10}$ rods of 25 mm and 20 mm in diameter, respectively, are shown in Fig. 6.

Figure 7 shows the effect of the substituting content $x$ of La on the $d_c$, and the thermal parameters of $T_g/T_l$, $\gamma$, $\Delta T_l$ and $\Delta T_x$ for the $(La_xCe_{1-x})_{65}Al_{10}Co_{25}$ alloys. With the increase of substituting content of La, the change of $d_c$, characteristic of the GFA of alloys, shows an asymmetrically pyramidic curve. When the $x$ reaches to 0.7, the maximal $d_c$ for the GFA of $(La_{0.7}Ce_{0.3})_{65}Al_{10}Co_{25}$ alloy is 25 mm. The coexistence of La and Ce in the alloys also induces a larger nominal supercooled degree, $\Delta T_l$, above 40 K with the substituting content $x$ from 0.3 to 0.8, consistent with the composition range with high GFA identified by $d_c$. The change of $\Delta T_x$ is similar with that of $d_c$ except for the excursion of their maximal values. The $T_g/T_l$, $T_g/T_{l'}$ (not shown in Fig. 7) and $\gamma$ show maximal values at the substituting La content of 0.5.

In a wide range of substitution composition ($0.4 \leq x \leq 0.8$), the glassy samples with a diameter above 10 mm can be fabricated, while the value of $d_c$ for the corresponding single lanthanide based alloys, $La_{65}Al_{10}Co_{25}$ and $Ce_{65}Al_{10}Co_{25}$, is only 2 mm. In order to further confirm that the coexistence of similar atoms of La and Ce plays a key role to improve the GFA in (La-Ce)-Al-Co system, the GFA of La-Al-Co and Ce-Al-Co systems were also evaluated in the corresponding composition maps, shown in Fig. 8. Although the compositions with the best GFA for La-Al-Co (Fig. 8 (a)) and Ce-Al-Co (Fig. 8 (b)) are excursive from the optimized one for (La-Ce)-Al-Co, the maximal $d_c$ in La-Al-Co and Ce-Al-Co systems is not more than 6 mm, far smaller than 25 mm of $(La_{0.7}Ce_{0.3})_{65}Al_{10}Co_{25}$. It strongly supports that the similar atom coexistence of La and Ce can significantly improve the GFA of (La-Ce)-Al-Co BMGs.

Because La (atomic number: 57) and Ce (atomic number: 58) lie in the neighboring positions in the PTE and both belong to the lanthanide elements, the difference of atomic size of La (187 pm) and Ce (182 pm) is small (only 2.7 %) [37], and the chemical properties of La and Ce are similar, characteristic of the heat of mixing of La-Ce pair is 0 kJ/mol [38]. So the mismatch of atomic sizes and the negative heats of mixing, mentioned above, can't give a reasonable explanation for the improvement of GFA by the substitution of La for Ce in the (La-Ce)-Al-Co system. In the following section, we will perform a further particular thermodynamic discussion and give a possible explanation for the effect of the row substitution with similar atoms on GFA.

3.3 Thermodynamic Analysis

Above experiments have demonstrated that the partial substitution of La for solvent Ce can significantly improve the GFA of $(La_xCe_{1-x})_{65}Al_{10}Co_{25}$ alloys. When the similar atoms, La and Ce, coexist in an appropriate composition ratio, the optimal composition with the best GFA can be obtained. To explain the remarkable improvement of GFA induced by the coexistence of similar atoms, we evaluated the mixing Gibbs free energy ($\Delta G$) in our alloy systems.

From the viewpoint of thermodynamics, the GFA of BMGs, to a certain extent, has a connection with the difference in energy between the solid glass and its liquid state, $G^s - G^l$, which reflects the stability of glassy state. It has been assumed that $G^s - G^l$ is proportional to the free energy of mixing $\Delta G$ of the liquid phase [39], so it is reasonable to understand the substitution effect on GFA by comparing the mixing Gibbs free energy of ante- and post-substitution ($\Delta G_a$ and $\Delta G_p$). $\Delta G$ of multicomponent system is defined as follows for standard states:

$$\Delta G = \Delta H - T\Delta S \qquad (1)$$

where $\Delta H$ is the increment of mixing enthalpy, $\Delta S$ is the increment of mixing entropy and $T$ is the temperature.

Because the accurate evaluation of thermodynamic parameters for glassy system is very difficult, some assumptions must be proposed. Considering the similarity between La and Ce as solvent atoms in the row substitution, according to Miracle's structural model for metallic glasses [37,40], we consider that this partial substitution would not cause a significant change of glassy structure. It means that the substituting atoms La will lie on the positions which are previously occupied by the substituted atoms Ce in the glassy structure, like the atomic replacement in an ideal substitutional solid solution, so the coordination number and the kinds of coordinated elements are not significantly changed. In our study, the glassy structure is approximatively regarded as a structure of ideal multicomponent solid solution without long range periodicity. $\Delta S$ is calculated as the sum of the configurational entropy ($\Delta S_{conf}$) and the mismatch entropy ($\Delta S_{mis}$) which is a function of the mismatch of component atomic size and caused by the structure modification from the periodic ideal solid solution to nonperiodic glassy structure, that is:

$$\Delta S = \Delta S_{conf} + \Delta S_{mis} \qquad (2)$$

For the ideal multicomponent solid solution consisting of N elements, the $\Delta S_{conf}$ can be expressed as equation (3):

$$\Delta S_{conf} = -R \sum_{i=1}^{N} (c_i \ln c_i) \qquad (3)$$

where R is the gas constant, $c_i$ is the mole concentration for i element. $\Delta H$ can be calculated as equation (4):

$$\Delta H = \sum_{i=1, i \neq j}^{N} \Omega_{ij} c_i c_j \qquad (4)$$

where $\Omega_{ij}$ is the interaction parameter of ideal solution between i and j element, related with the coordination numbers.

For simplicity, we adopted the analysis method by Takeuchi *et al.* [41] to deduce the thermal parameters, $\Delta H$ and $\Delta S_{mis}$ ($\Delta S_{mis}$ was approximately replaced by the excess entropy of an ideal gas at the same pressure according to Takeuchi's deduction). Based on the above assumption for atomic substitution, we can calculate the $\Delta G$ of pseudo-ternary metallic glass at liquid state for a certain composition, and further compare the change of $\Delta G$ for ante- and post-substituted alloys. $\Omega_{ij}$ can be approximately substituted with the relation equation:

$$\Omega_{ij} = 4\Delta H_{ij}^{mix} \tag{5}$$

where $\Delta H_{ij}^{mix}$ is the enthalpy of mixing which is calculated in a binary i-j system at the equiatomic composition [38]. $\Delta S_{mis}$ can be treated as the discussions in references [40].

Figure 9 shows the effect of substitution content $x$ of La on $\Delta G$ for the $(La_xCe_{1-x})_{65}Al_{10}Co_{25}$ BMGs. The temperature of system was chosen at 1050 K above the liquid temperature of all alloys. The critical diameter of $(La_xCe_{1-x})_{65}Al_{10}Co_{25}$ BMGs, characteristic of the GFA, is also given in Fig. 9 for comparison. We noticed that because the enthalpy of mixing ($\Delta H^{mix}$) between La and Ce is 0 kJ/mol, and there is no considerable difference in $\Delta H^{mix}$ between the atomic pairs of La-Co (-17 kJ/mol) and Ce-Co (-18 kJ/mol) as well as between those of La-Al (-38 kJ/mol) and Ce-Al (-38 kJ/mol) [40], the change of $\Delta H$ among the different substituted alloys is indistinctive. Furthermore, because of the similar atomic size between La and Ce, the change of the mismatch entropy, resulting from atomic size mismatch, will also be little. So the main difference in $\Delta G$ among the substituted alloys is mainly attributed to the change of $\Delta S_{conf}$. The $\Delta G$ reduces steeply with the partial substitution of La for Ce (or of Ce for La) in Ce-Al-Co (or La-Al-Co) alloy, which implies that the coexistence of La and Ce can make the system more stable and obviously improve the GFA of the resulting alloy comparing with the single lanthanide based alloys. The minimal value of $\Delta G$, corresponding to the most stable composition for the glassy alloys by thermodynamic analysis, appears at the equiatomic ratio for La and Ce. The experimental results of the $d_c$ are fundamentally consistent with the calculation of $\Delta G$ except the deviation of the best composition between the experiments ($(La_{0.7}Ce_{0.3})_{65}Al_{10}Co_{25}$) and the thermodynamic analysis ($(La_{0.5}Ce_{0.5})_{65}Al_{10}Co_{25}$). The reason for this deviation maybe that the dynamic process for a competition process of every resulting alloy after the row substitution between supercooled liquid and crystalline phases was not considered in this model. Nevertheless, this thermodynamic analysis can give a reasonable explanation for the obvious improvement of GFA caused by the row substitution, while the conventional criteria, such as atomic size mismatch and negative heats of mixing, are incompetence.

## 4. Conclusions

In this paper, the glass-formation range of BMGs based on lanthanum and cerium was pinpointed in pseudo-ternary (La-Ce)-Al-Co composition map by copper mold casting. The influence of the substitution of La for solvent Ce in $(La_xCe_{1-x})_{65}Al_{10}Co_{25}$ system on the GFA was evaluated in detail. The $(La_{0.7}Ce_{0.3})_{65}Al_{10}Co_{25}$ alloy with superior GFA, identified by forming the fully glassy rod of 25 mm in diameter by tilt-pour casting, was found out through this row substitution. Because this substitution with the similar atoms of La and Ce can't be explained by former criteria of GFA, e.g. atomic size mismatch and negative heats of mixing, a thermodynamic model was proposed to evaluate the effect of the coexistence of La and Ce in (La-Ce)-Al-Co system. The decrease of $\Delta G$ caused by the configurational entropy can give a reasonable explanation for the obvious improvement of GFA by this substitution.

## Acknowledgements

This work was financially supported by National Nature Science Foundation of China (No. 50225103 and No. 50471001).

**Figure Captions**

Fig. 1 The composition map for the glass-forming range in the $(La_{0.5}Ce_{0.5})$-Al-Co system. The symbols represent: ▲ BMG; ◐ crystalline.

Fig. 2 XRD patterns of the as-cast rods of 12 mm in diameter for $(La_{0.5}Ce_{0.5})_{65}Al_{10}Co_{25}$, $(La_{0.5}Ce_{0.5})_{65}Al_{15}Co_{20}$ and $(La_{0.5}Ce_{0.5})_{70}Al_{15}Co_{15}$ BMGs.

Fig. 3 The 3D contour map of liquidus surface for $(La_{0.5}Ce_{0.5})$-Al-Co alloys. The eutectic composition and the composition range for 12 mm BMGs are marked.

Fig. 4 DSC curves of the 2 mm glassy rods of $(La_xCe_{1-x})_{65}Al_{10}Co_{25}$ alloys ($0 < x < 1$) at the rate of 0.33 K/s for: (a) heating procedure; (b) cooling procedure.

Fig. 5 XRD patterns for the $(La_xCe_{1-x})_{65}Al_{10}Co_{25}$ glassy rods ($0 < x < 1$) with the maximal size in diameter.

Fig. 6 The outer shape and surface appearance of the as-cast $(La_{0.7}Ce_{0.3})_{65}Al_{10}Co_{25}$ and $(La_{0.6}Ce_{0.4})_{65}Al_{10}Co_{25}$ rods of 25 mm and 20 mm in diameter, respectively.

Fig. 7 The effect of the substituting content $x$ of La on $d_c$, $T_g/T_l$, $\gamma$, $\Delta T_l$ and $\Delta T_x$ for the $(La_xCe_{1-x})_{65}Al_{10}Co_{25}$ BMGs.

Fig. 8 The composition map for the glass-forming range in the La-Al-Co and Ce-Al-Co systems. The symbols represent: ▲ BMGs with $d_c$ of 2 mm; ● BMGs with $d_c$ of 4 mm; ■ BMGs with $d_c$ of 6 mm; ◐ crystalline.

Fig. 9 The effect of substitution content $x$ of La on the mixing Gibbs free energy, $\Delta G$, for the $(La_xCe_{1-x})_{65}Al_{10}Co_{25}$ BMGs.

**Figure 1 Glass Forming Range of (La-Ce)-Al-Co**
Click here to download high resolution image

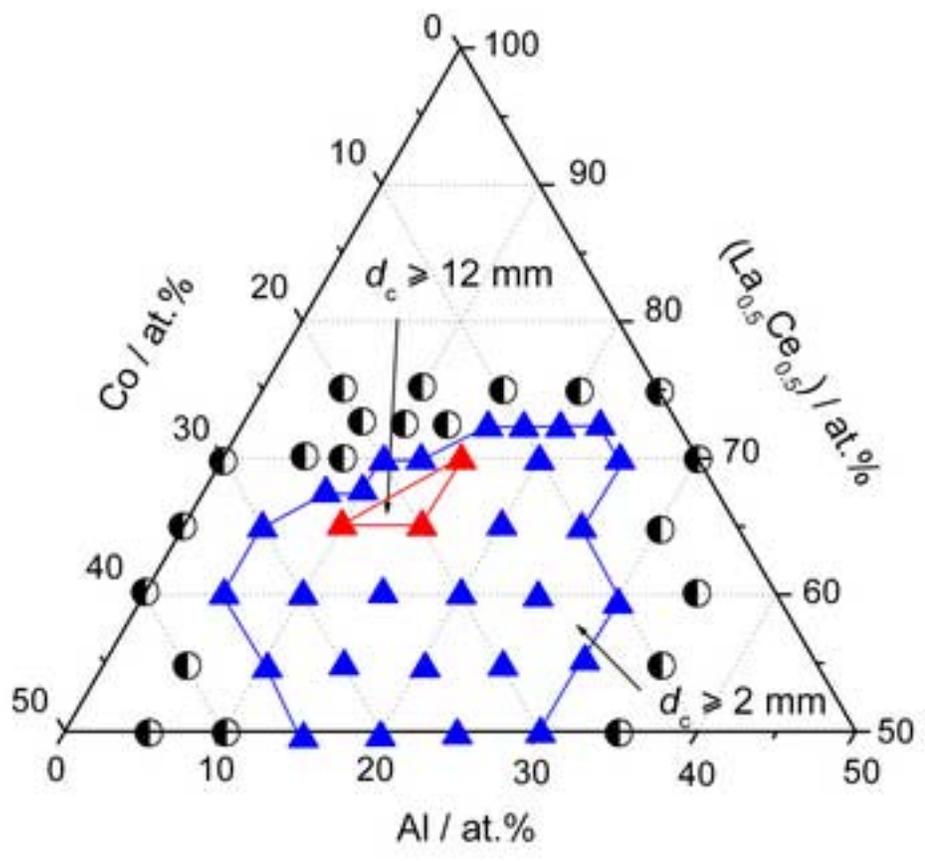

**Figure 2 XRD Patents of (La-Ce)-Al-Co**
[Click here to download high resolution image]

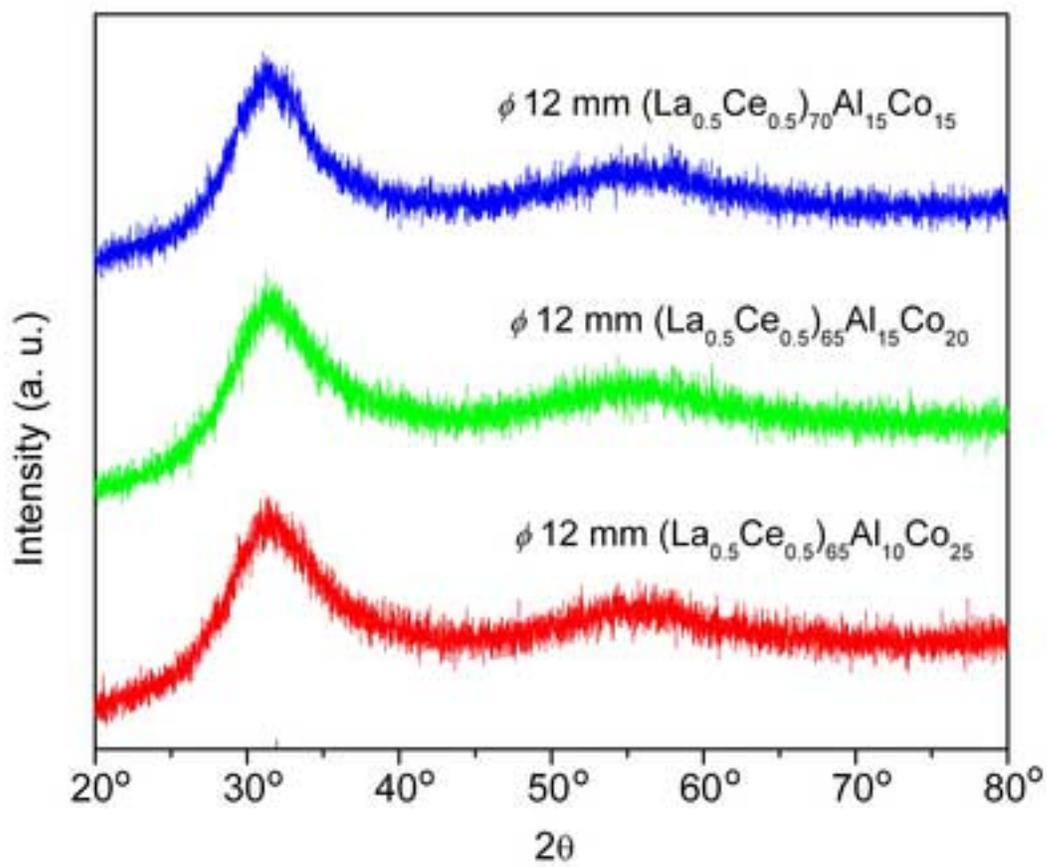

**Figure 3 3D-TI**
Click here to download high resolution image

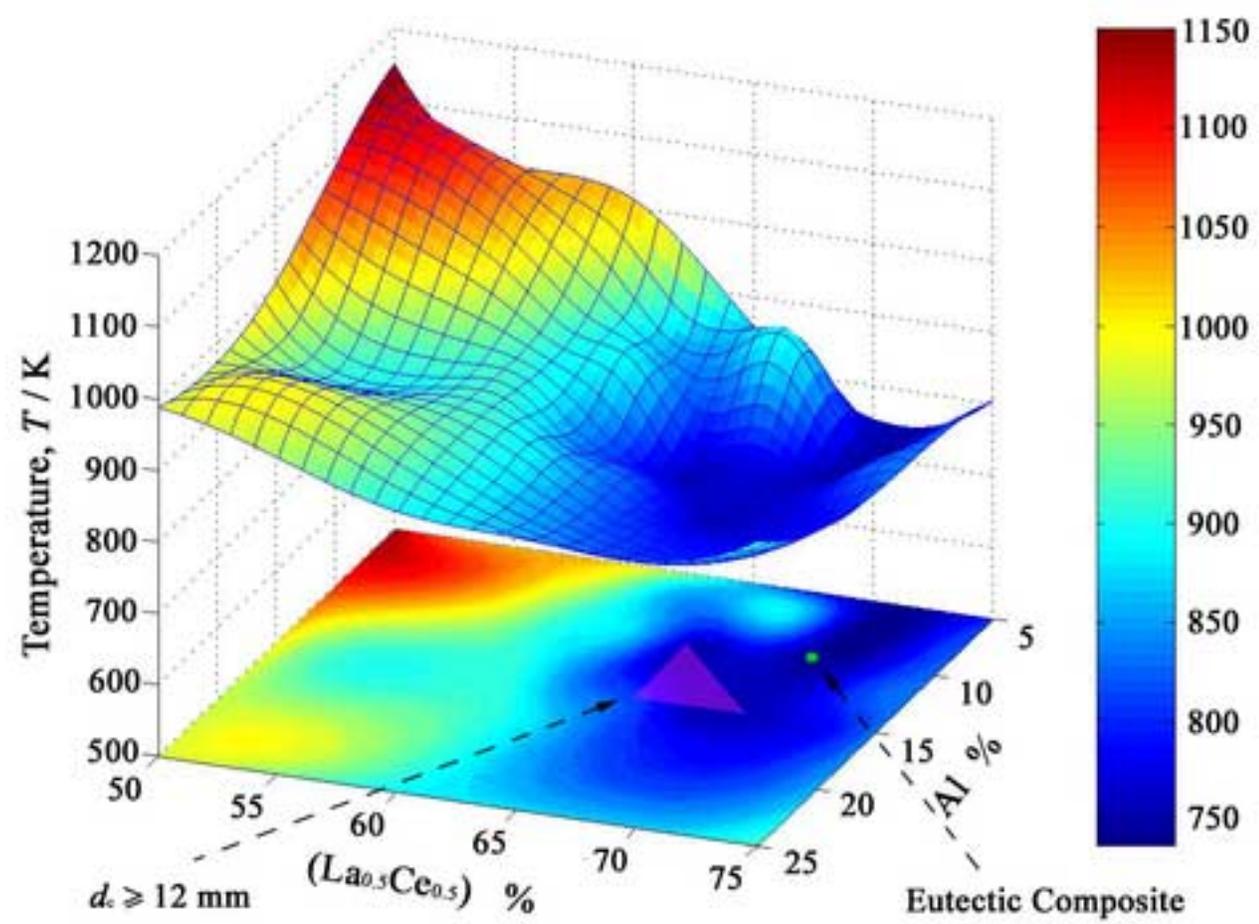

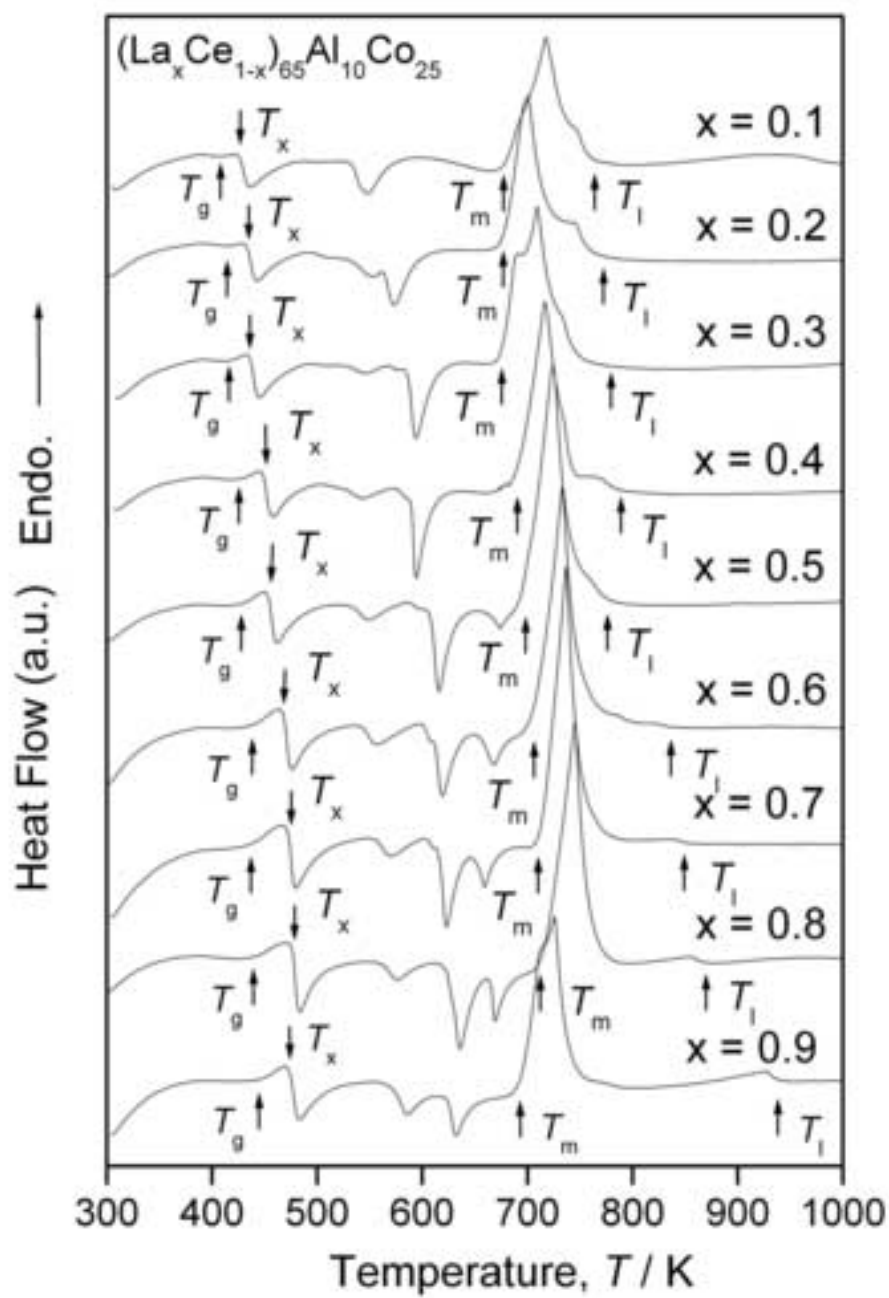

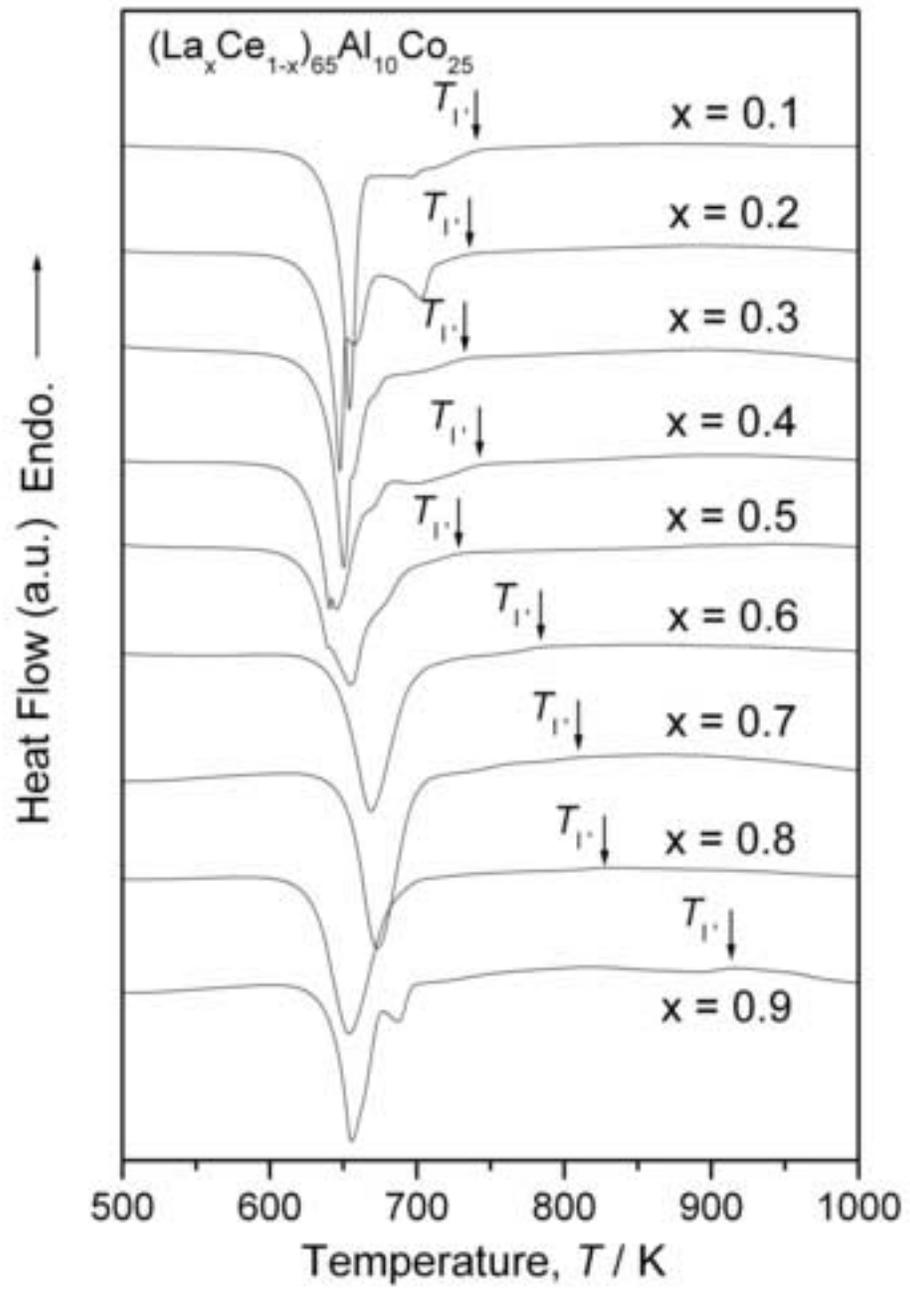

**Figure 5** XRD Patents of (LaxCe1-x)-Al-Co
Click here to download high resolution image

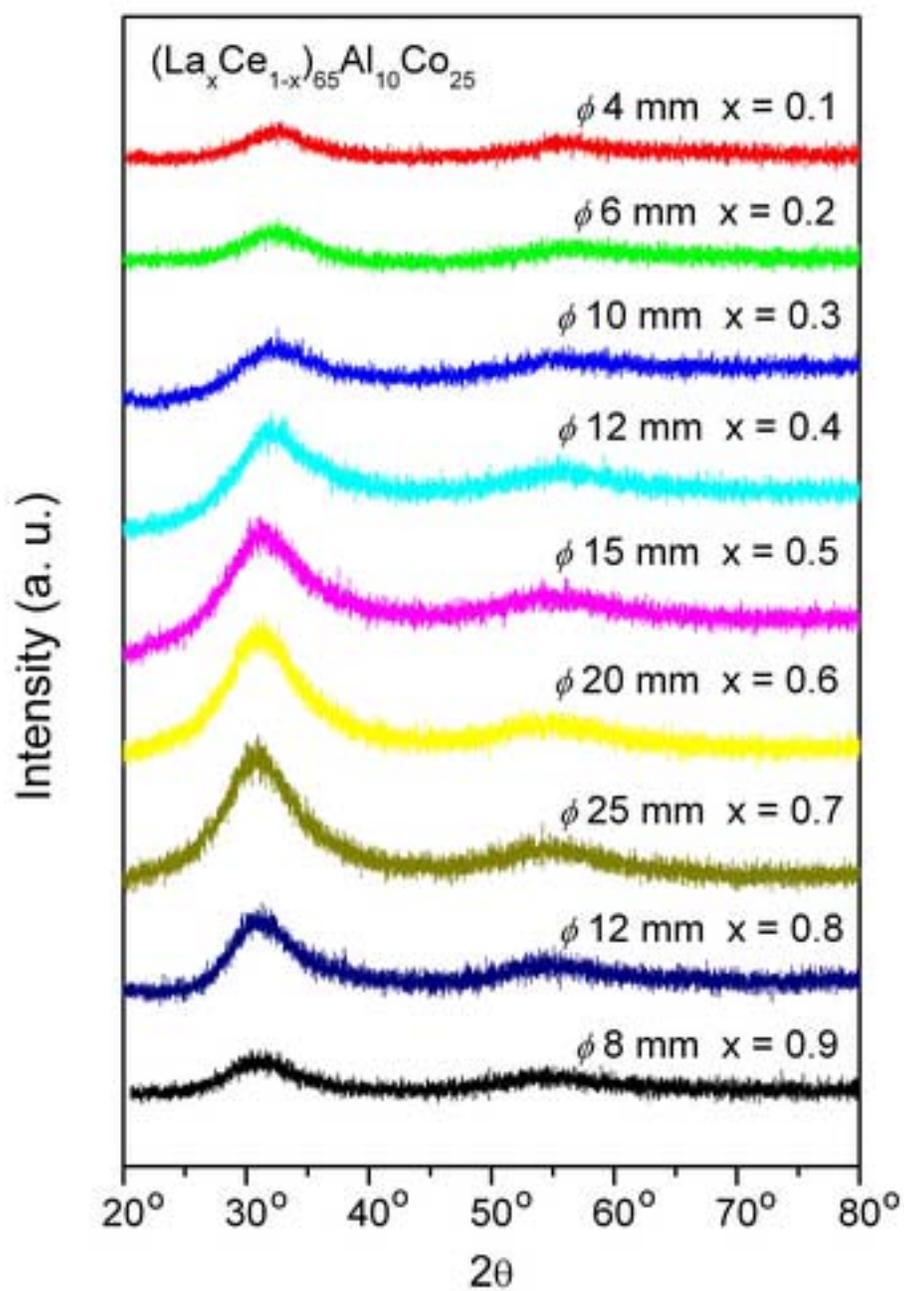

**Figure 6 La-Ce-Al-Co-25,20mm**
**Click here to download high resolution image**

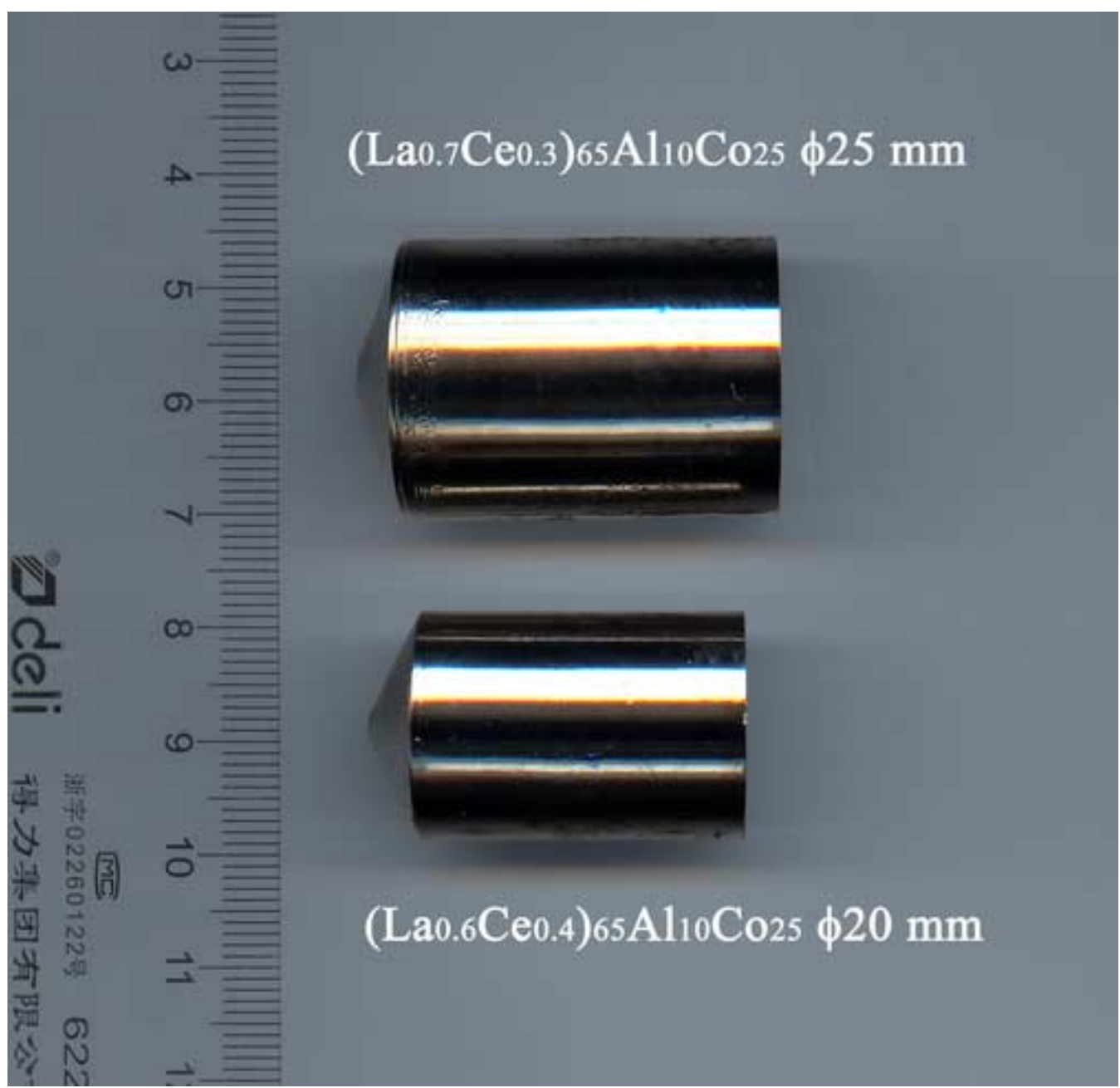

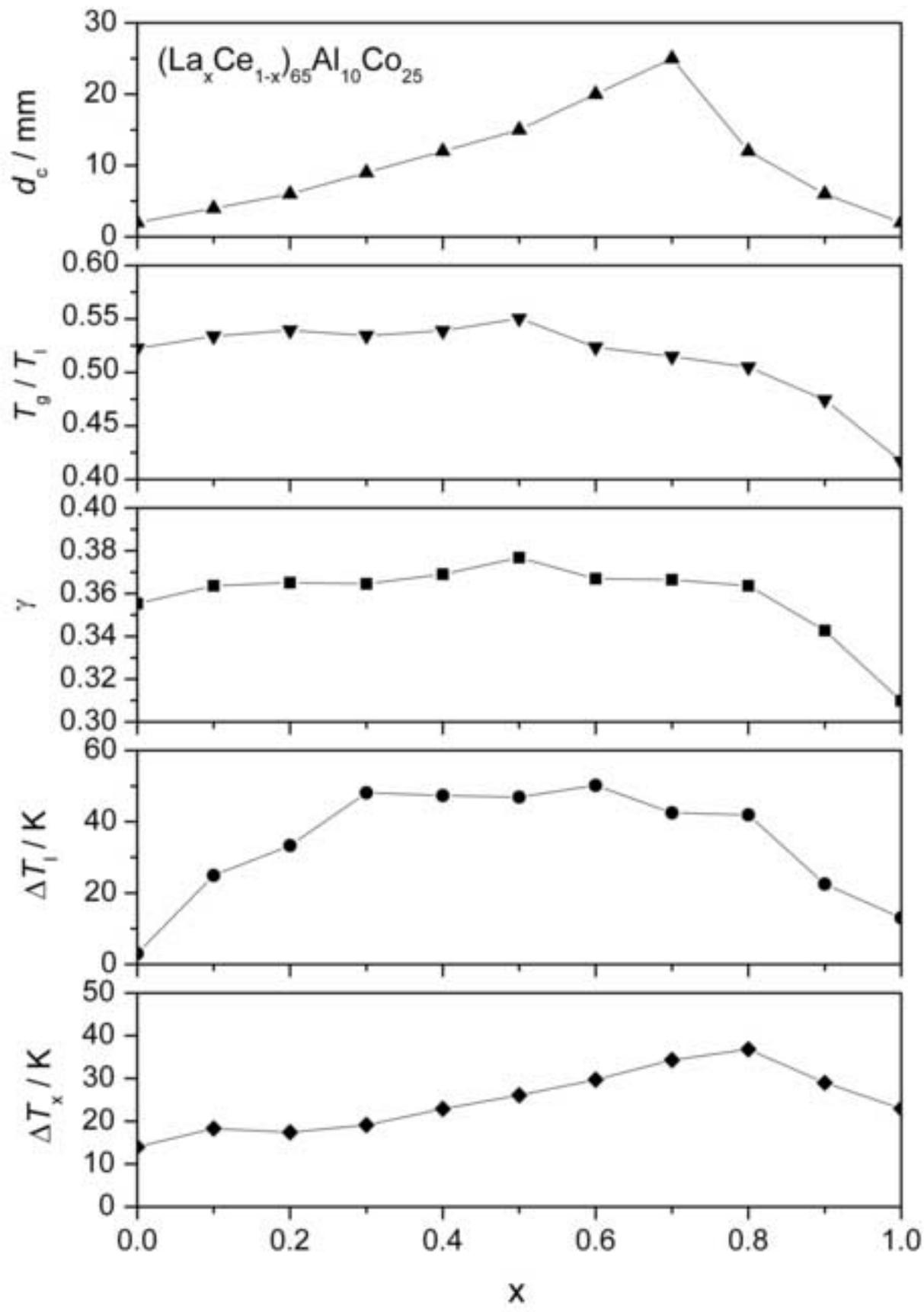

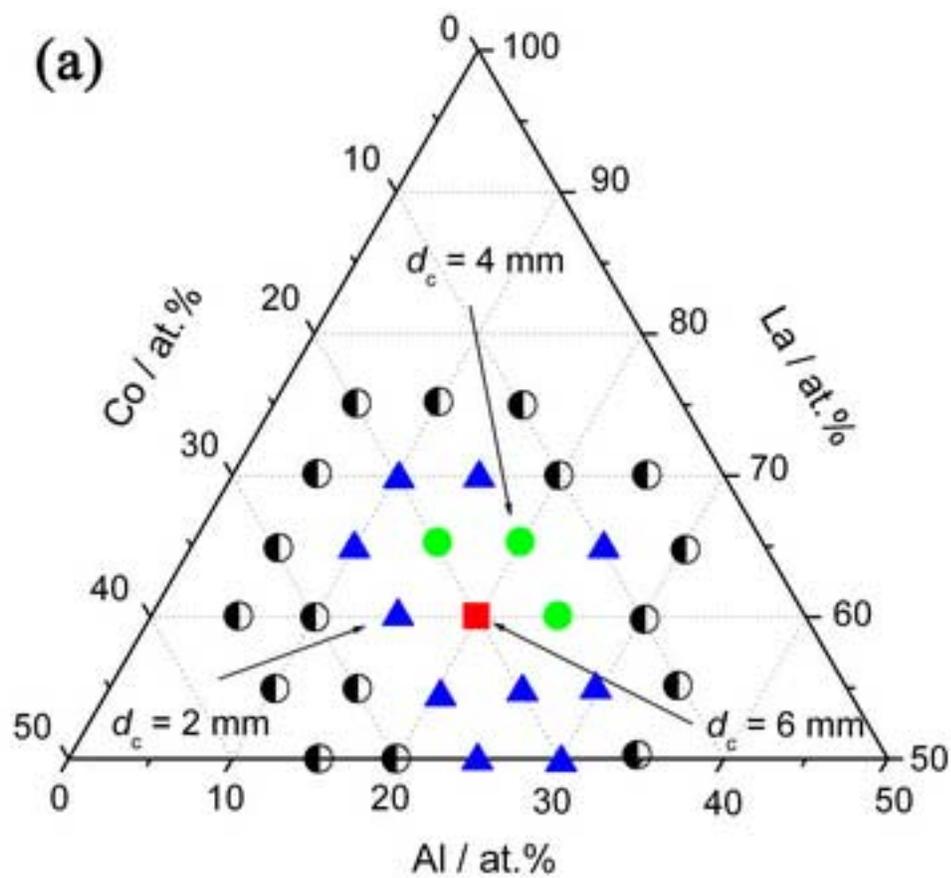
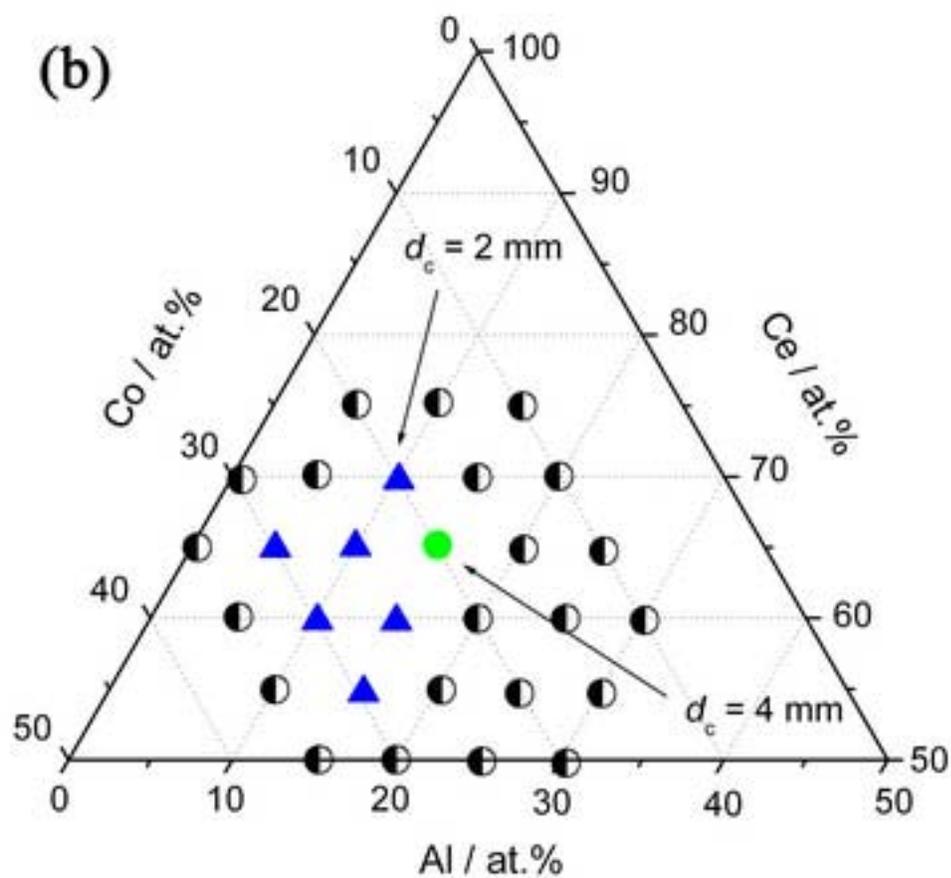

**Figure 9 Gibbs Free engery**
**Click here to download high resolution image**

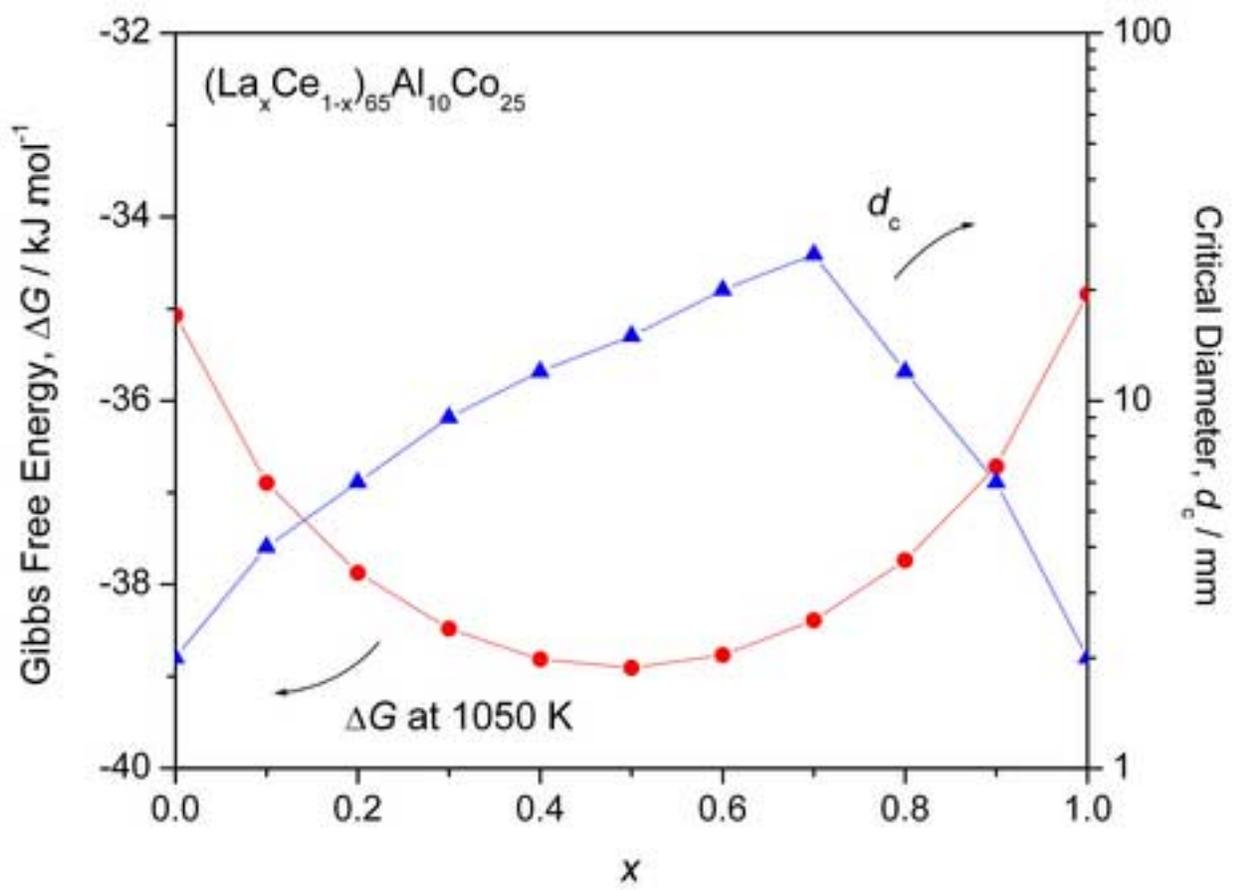

Table I Thermal parameters of $(La_{0.5}Ce_{0.5})$-Al-Co BMGs.

| Alloy | $T_g$ / K | $T_x$ / K | $T_m$ / K | $T_l$ / K | $\Delta T_x$ / K | $T_g/T_l$ | $\gamma$ |
|---|---|---|---|---|---|---|---|
| $(La_{0.5}Ce_{0.5})_{70}Al_{10}Co_{20}$ | 414 | 434 | 698 | 744 | 20 | 0.56 | 0.375 |
| $(La_{0.5}Ce_{0.5})_{70}Al_{15}Co_{15}$ | 421 | 445 | 713 | 764 | 24 | 0.55 | 0.376 |
| $(La_{0.5}Ce_{0.5})_{70}Al_{20}Co_{10}$ | 429 | 483 | 722 | 798 | 54 | 0.54 | 0.394 |
| $(La_{0.5}Ce_{0.5})_{70}Al_{25}Co_{5}$ | 461 | 484 | 716 | 874 | 23 | 0.53 | 0.363 |
| $(La_{0.5}Ce_{0.5})_{65}Al_{5}Co_{30}$ | 402 | 439 | 696 | 858 | 37 | 0.47 | 0.348 |
| $(La_{0.5}Ce_{0.5})_{65}Al_{10}Co_{25}$ | 427 | 453 | 696 | 776 | 26 | 0.55 | 0.376 |
| $(La_{0.5}Ce_{0.5})_{65}Al_{15}Co_{20}$ | 440 | 502 | 694 | 812 | 62 | 0.54 | 0.400 |
| $(La_{0.5}Ce_{0.5})_{65}Al_{20}Co_{15}$ | 444 | 516 | 716 | 831 | 72 | 0.53 | 0.404 |
| $(La_{0.5}Ce_{0.5})_{65}Al_{25}Co_{10}$ | 462 | 502 | 719 | 880 | 40 | 0.52 | 0.374 |
| $(La_{0.5}Ce_{0.5})_{60}Al_{5}Co_{35}$ | 436 | 455 | 694 | 1010 | 19 | 0.43 | 0.315 |
| $(La_{0.5}Ce_{0.5})_{60}Al_{10}Co_{30}$ | 448 | 478 | 695 | 902 | 30 | 0.50 | 0.354 |
| $(La_{0.5}Ce_{0.5})_{60}Al_{15}Co_{25}$ | 459 | 545 | 702 | 904 | 86 | 0.51 | 0.400 |
| $(La_{0.5}Ce_{0.5})_{60}Al_{20}Co_{20}$ | 475 | 554 | 693 | 916 | 79 | 0.52 | 0.398 |
| $(La_{0.5}Ce_{0.5})_{60}Al_{25}Co_{15}$ | 477 | 522 | 715 | 892 | 45 | 0.53 | 0.381 |
| $(La_{0.5}Ce_{0.5})_{55}Al_{10}Co_{35}$ | 461 | 514 | 713 | 1021 | 53 | 0.45 | 0.347 |
| $(La_{0.5}Ce_{0.5})_{55}Al_{15}Co_{30}$ | 475 | 556 | 691 | 902 | 81 | 0.53 | 0.404 |
| $(La_{0.5}Ce_{0.5})_{55}Al_{20}Co_{25}$ | 498 | 561 | 725 | 961 | 63 | 0.52 | 0.384 |
| $(La_{0.5}Ce_{0.5})_{55}Al_{25}Co_{20}$ | 496 | 561 | 717 | 942 | 65 | 0.53 | 0.390 |
| $(La_{0.5}Ce_{0.5})_{50}Co_{35}Al_{15}$ | 492 | 562 | 715 | 1005 | 70 | 0.49 | 0.375 |
| $(La_{0.5}Ce_{0.5})_{50}Co_{30}Al_{20}$ | 506 | 568 | 711 | 971 | 62 | 0.52 | 0.385 |

Table II Critical diameter ($d_c$), thermal parameters and density ($\rho$) of the ($La_xCe_{1-x}$)-Al-Co BMGs and the corresponding La- and Ce-based BMGs [24,25].

| Alloys | $d_c$ / mm | $T_g$ / K | $T_x$ / K | $T_m$ / K | $T_l$ / K | $T_{l'}$ / K | $\Delta T_x$ / K | $\Delta T_l$ / K | $T_g/T_l$ | $T_g/T_{l'}$ | $\gamma$ | $\rho$ / g·cm$^{-3}$ |
|---|---|---|---|---|---|---|---|---|---|---|---|---|
| $Ce_{65}Al_{10}Co_{25}$ | 2 | 396 | 410 | 676 | 758 | 755 | 14 | 3 | 0.52 | 0.52 | 0.355 | 7.22 |
| $(La_{0.1}Ce_{0.9})_{65}Al_{10}Co_{25}$ | 4 | 408 | 426 | 677 | 764 | 740 | 18 | 24 | 0.53 | 0.55 | 0.363 | 7.18 |
| $(La_{0.2}Ce_{0.8})_{65}Al_{10}Co_{25}$ | 6 | 414 | 432 | 676 | 770 | 736 | 18 | 34 | 0.54 | 0.56 | 0.365 | 7.12 |
| $(La_{0.3}Ce_{0.7})_{65}Al_{10}Co_{25}$ | 9 | 416 | 436 | 675 | 778 | 731 | 20 | 47 | 0.54 | 0.57 | 0.365 | 7.00 |
| $(La_{0.4}Ce_{0.6})_{65}Al_{10}Co_{25}$ | 12 | 425 | 448 | 689 | 789 | 742 | 23 | 47 | 0.54 | 0.57 | 0.369 | 6.96 |
| $(La_{0.5}Ce_{0.5})_{65}Al_{10}Co_{25}$ | 15 | 427 | 453 | 696 | 776 | 729 | 26 | 47 | 0.55 | 0.58 | 0.376 | 6.81 |
| $(La_{0.6}Ce_{0.4})_{65}Al_{10}Co_{25}$ | 20 | 437 | 467 | 705 | 835 | 785 | 30 | 50 | 0.52 | 0.56 | 0.367 | 6.70 |
| $(La_{0.7}Ce_{0.3})_{65}Al_{10}Co_{25}$ | 25 | 437 | 472 | 710 | 850 | 807 | 35 | 43 | 0.51 | 0.54 | 0.367 | 6.59 |
| $(La_{0.8}Ce_{0.2})_{65}Al_{10}Co_{25}$ | 12 | 439 | 476 | 712 | 869 | 827 | 37 | 42 | 0.50 | 0.53 | 0.364 | 6.49 |
| $(La_{0.9}Ce_{0.1})_{65}Al_{10}Co_{25}$ | 6 | 444 | 474 | 694 | 937 | 914 | 30 | 23 | 0.47 | 0.48 | 0.343 | 6.41 |
| $La_{65}Al_{10}Co_{25}$ | 2 | 434 | 457 | 694 | 1041 | 1028 | 23 | 13 | 0.42 | 0.42 | 0.310 | 6.32 |
| $La_{62}Al_{15.7}(Ni_{0.5}Co_{0.5})_{22.3}$ | 11 | 422 | 460 | 675 | 722 | - | 38 | - | 0.62 | 0.58 | 0.402 | - |
| $Ce_{70}Al_{10}Cu_{20}$ | 2 | 341 | 408 | 647 | 722 | - | 67 | - | 0.53 | 0.47 | 0.383 | 6.699 |
| $Ce_{68}Al_{10}Cu_{20}Nb_2$ | 8 | 345 | 421 | 646 | 721 | - | 76 | - | 0.53 | 0.48 | 0.395 | 6.738 |
| $Ce_{68}Al_{10}Cu_{20}Co_2$ | 10 | 352 | 419 | 615 | 716 | - | 67 | - | 0.57 | 0.49 | 0.392 | 6.752 |